\documentclass[11pt]{amsart}

\usepackage{amsmath}
\usepackage{amsfonts}
\usepackage{amssymb}
\usepackage{latexsym}

 \advance\hoffset by -0.4 truecm

\newtheorem{definition}{Definition}
\newtheorem{theorem}{Theorem}
\newtheorem{corollary}{Corollary}
\newtheorem{remark}{Remark}

\newcommand{\R}{\text{\rm Re }}
\newcommand{\I}{\text{\rm Im }}

\title{Univalent functions and integrable systems}
\author{Dmitri Prokhorov and Alexander Vasil'ev}

\address{Department of Mathematics and Mechanics, Saratov State University, Saratov 410026,
Russia}

\email{ProkhorovDV@info.sgu.ru}

\address{Matematisk institutt, Universitetet i Bergen, Johannes Brunsgate 12, N-5008 Bergen, Norway}
\email{alexander.vasiliev@uib.no}



\begin{document}

\begin{abstract}
We study one-parameter expanding evolution families of simply
connected domains in the complex plane described by infinite systems
of evolution parameters. These evolution parameters in some cases
admit Hamiltonian formulation and lead to integrable systems. One
example of such parameters is complex moments for the Laplacian
growth that form a Whitham-Toda integrable hierarchy. Another
example we deal with is related to expanding coefficient bodies for
conformal maps given by L\"owner subordination chains. The
coefficients bodies are proved to form a Liouville partially
integrable Hamiltonian system for each fixed index and the first
integrals are obtained. We also discuss the contact structure of
this system.
\end{abstract}

\maketitle

\section{Introduction}
Complex dynamical systems, iterations and construction of Lie semigroups with respect to the composition operation
lead to the study of expanding systems of plane domains. These systems can be thought of as one-parameter families of
domains (typically simply connected and univalent) $\Omega(t)$ that form {\it subordination chains}  in the Riemann sphere $\hat{\mathbb
C}$, which are defined for $0\leq t< T$ where $T$ may be $\infty$. This means that
$\Omega(t)\subset \Omega(s)$,  $\Omega(t)\neq \Omega(s)$, whenever $t<s$.

 Kufarev
\cite{Kufarev} studied a one-parameter family of domains
$\Omega(t)$, and regular functions $g(z,t)$, $g:\,\Omega(t)\times [0,T)\to U$,
where $U=\{\zeta:\, |\zeta|<1\}$. He proved differentiability of $g(z,t)$ with respect
to $t$ for $z$ from the Carath\'eodory kernel $\Omega(0)$ of
$\Omega(t)$.

Sometimes, the evolution of $\Omega(t)$ can be characterized by
infinite sets of {\it evolution parameters} given as a result of
integration of certain dynamical systems. Hamiltonian interpretation
of such systems yields their integrability. The interest to
integrable systems is caused first of all by Hamiltonian systems of
mathematical physics where the Hamiltonian is thought of as an
energy functional and its critical  points correspond to minimal
energy of particle systems. However, other areas of mathematics also
use Hamiltonian formalism, e.g., optimal control, classical
mechanics, etc.

One of the examples of integrable systems of evolution parameters
corresponding to one-parameter families of expanding domains is the
problem of Laplacian growth (Hele-Shaw problem). The Laplacian
growth means the Dirichlet problem for a harmonic potential where
the boundary of the phase domain $\Omega(t)$ is unknown a-priori
(free boundary), and in fact, is defined by the normality of its
motion. The classical (strong) solvability of this problem suggests
that phase domains $\Omega(t)$ are bounded by analytic curves. It
turns out, that given an initial domain $\Omega(0)$ the set of
Richardson's moments (see the definition in \cite{Richardson})
solves completely the inverse potential problem, thus describing the
evolution of phase domains as long as the classical solution exists.
As it has been shown in \cite{Agam}, \cite{Kostov},
\cite{Marshakov}, \cite{Wiegmann}, the Laplacian growth problem can
be embedded into a larger hierarchy of domain variations
(Whitham-Toda hierarchy) for which all the complex moments are
treated as independent variables (generalized time variables), and
form an integrable system.

Another example of evolution parameters is the set of coefficients
of the Riemann maps of $U$ onto the family of the phase domains
$\Omega(t)$ that has attracted attention of the complex analysts for
a more than eighty year period of the last century. The Bieberbach
conjecture \cite{Bieberbach} has proved to be the most intriguing
problem that forced investigations in geometric function theory
during this period. It finally has been solved in 1984 by de~Branges
\cite{Branges}, who proved that the $n$-th Taylor coefficient $b_n$
of a conformal homeomorphism $f(\zeta)$, normalized by $f(0)=0$,
$f'(0)=1$, does not exceed its index: $|b_n|\leq n$. This class of
functions we denote by $S$. The inequality is sharp and the equality
is attained only for rotations of the Koebe function
$k(z)=z(1-z)^{-2}$. De~Branges' proof has put an end to the numerous
attempts to prove or disprove the Bieberbach conjecture. However, a
much more general and difficult problem is to describe the range of
$n$ first coefficients as a set of values $V_n=\{(b_2,\dots,
b_n),\,f\in S\}$. In the trivial case $V_2$ is the disk of radius 2.
Only the first non-trivial coefficient body $V_3=(b_2,b_3)$ has been
described completely by Schaeffer and Spenser in 1951 in their
famous monograph \cite{SS}. A qualitative description of $V_n$,
$n\geq 3$, has been partially given in \cite{Babenko}. Apart from
these two monographs there are only few works where a progress in
such a problem has been made (see, e.g., \cite{Prokhorov}).

The key stages of our presentation are as follows. Firstly, a
boundary point of $V_n$ is given by a unique function whose first
coefficients $b_2,\dots, b_n$ define the rest of coefficients.
Secondly, this function unlikely the Laplacian growth, presents a
full mapping (the complement of the image of $U$ is of measure zero)
of the unit disk onto the complex plane minus an analytic graph with
the root at infinity. Thirdly, erasing this graph we obtain a family
of expanding domains completely described by the initial domain, and
in its turn by the point of $\partial V_n$. Finally, we obtain a
$(2n-3)$ dimensional family of evolution dynamics completely
described as a flow generated by a certain vector field, which
admits a Hamiltonian formulation. It is proved to be partially
integrable in the Liouville sense and the first integrals are
obtained. Moreover, it turns out that the first integrals that
generate additional directions form a horizontal space so this
Hamiltonian system possesses a contact structure. It is important
that the evolution exists all the time $0\leq t <\infty$.

Our paper is devoted to a detail realization of this plan and the
description of this system of evolution parameters, its Hamiltonian
formulation, and finally, its integration. The central idea of the
Hamiltonian formulation is forced by the L\"owner evolution (see,
e.g., \cite{Aleksandrov, Loewner, Pom1}) that is based on evolution
equations in partial and ordinary derivatives mutually linked. The
controllable dynamical system for evolution parameters turns out to
be Hamiltonian under the assumption that the necessary optimality
conditions hold. So, we are especially interested in subordination
chains of domains generated by Riemann maps corresponding to the
critical controls, and therefore, to the trajectories, that in
particular, give all boundary points to the coefficient body for
univalent maps.

From one side our main results  describe evolution coefficient
systems for certain type of L\"owner's dynamics of domains and maps
that admit the Hamiltonian formulation and are integrable. The
coefficient bodies will form a certain type of hierarchy (dependent
on the coefficient indices) and the first integrals will be derived.
From the other side, we give all boundary points of the coefficient
bodies $V_n$.

\section{The L\"owner and L\"owner-Kufarev evolution}

There are many papers and monographs devoted to the L\"owner and
L\"owner-Kufarev equations. However we revisit this theory from the
point of view of correspondence between L\"owner equations in
partial and ordinary derivatives and univalence of solutions to the
partial derivative L\"owner equation.

Let us consider a subordination chain  of simply connected
hyperbolic domains $\Omega(t)$ in the complex plane $\mathbb C$,
which is defined for $0\leq t< T$, where $T$ may be infinity. We
suppose that the origin is an interior point of the kernel of
$\{\Omega(t)\}_{t=0}^{T}$. Let us normalize the growth of the
evolution of this subordination chain by the conformal radius of
$\Omega(t)$ with respect to the origin to be $e^t$. By the Riemann
Mapping Theorem we construct a subordination chain of mappings
$f(\zeta,t)$, $\zeta\in U$,  where each function $\displaystyle
f(\zeta,t)=e^t\zeta+b_2(t)\zeta^2+\dots$ is a holomorphic univalent
map of $U$ onto $\Omega(t)$ for every fixed $t$. Pommerenke
\cite{Pom1, Pom2}  introduced such chains in order to generalize
L\"owner's equation. His result says that given a subordination
chain of domains $\Omega(t)$ defined for $t\in [0,T)$, there exists
an analytic regular function
$$p(\zeta,t)=1+p_1(t)\zeta+p_2(t)\zeta^2+\dots,\quad
\zeta\in U,$$ such that $\R p(\zeta, t)>0$  and
\begin{equation}
\frac{\partial f(\zeta,t)}{\partial t}=\zeta\frac{\partial
f(\zeta,t)}{\partial \zeta}p(\zeta,t),\label{LK}
\end{equation}
for $\zeta\in U$ and for almost all $t\in [0,T)$. The class of
holomorphic functions $p$ defined as above we denote by $C$. The
equation (\ref{LK}) is called the L\"owner-Kufarev equation due to
two seminal papers by L\"owner \cite{Loewner} with
\begin{equation}
p(\zeta,t)=\frac{e^{iu(t)}+\zeta}{e^{iu(t)}-\zeta},\label{yadro}
\end{equation}
where $u(t)$ is a continuous function regarding to $t\in [0,T)$,
 and by Kufarev \cite{Kufarev} in general case,
where this equation appeared for the first time.

The equation (\ref{LK}) represents a growing evolution of simply
connected domains. Let us consider the reverse process. Given an
initial domain $\Omega(0)\equiv \Omega_0$ (and therefore, the
initial mapping $f(\zeta,0)\equiv f_0(\zeta)$), and a function
$p(\zeta,t)$ from the class $C$, we solve the equation (\ref{LK})
and ask whether the solution $f(\zeta,t)$ represents a subordination
chain of simply connected domains. The initial condition
$f(\zeta,0)=f_0(\zeta)$ is not given on the characteristics of the
partial derivative equation (\ref{LK}), hence the solution exists
and is unique. Assuming $s$ as a parameter along the characteristics
we have $$ \frac{dt}{ds}=1,\quad \frac{d\zeta}{ds}=-\zeta
p(\zeta,t), \quad \frac{df}{ds}=0,$$ with the initial conditions
$t(0)=0$, $\zeta(0)=z$, $f(\zeta,0)=f_0(\zeta)$, where $z$ is in
$U$. This leads us to the Cauchy problem for the  L\"owner-Kufarev
equation in ordinary derivatives for a function $\zeta=w(z,t)$
\begin{equation}
\frac{dw}{dt}=-wp(w,t),\label{LKord}
\end{equation}
with the initial condition $w(z,0)=z$.

We see that the equation (\ref{LKord}) is exactly the characteristic
equation for (\ref{LK}). Unfortunately, this approach requires the
extension of $f_0(w^{-1}(\zeta,t))$ into $U$ because the solution to
(\ref{LK}) is the function $f(\zeta,t)$  given as
$f_0(w^{-1}(\zeta,t))$, where $\zeta=w(z,s)$ is a solution of the
initial value problem for the characteristic equation (\ref{LKord})
that maps $U$ into $U$. Therefore, {\bf the solution of the initial
value problem for the equation (\ref{LK}) may be non-univalent}.

Let $S$ stand for the usual class of all univalent holomorphic
functions $f(z)=z+a_2z^2+\dots$ in the unit disk. Solutions to the
equation (\ref{LKord}) are regular univalent functions
$w(z,t)=e^{-t}z+a_2(t)z^2+\dots$ in the unit disk that map $U$  into
itself. Conversely, every function from the class $S$ can be
represented by the limit
\begin{equation}
f(z)=\lim\limits_{t\to\infty}e^t w(z,t),\label{limit}
\end{equation}
where there exists a function $p(z,t)$ from the class $C$ for almost
all $t\geq 0$, such that $w(z,t)$ is a solution to the eqiation
(\ref{LKord}) (see \cite[pages 159--163]{Pom2}). Each function
$p(z,t)$ generates a unique function from the class $S$. The
reciprocal statement is not true. In general, a function $f\in S$
can be determined by different functions $p\in C$.

From \cite[page 163]{Pom2} it follows that we can guarantee the
univalence of the solutions to the L\"owner-Kufarev equation in
partial derivatives (\ref{LK}) assuming the initial condition
$f_0(\zeta)$ given by the limit (\ref{limit}) with the function
$p(\cdot,t)$ chosen to be the same in the equations (\ref{LK}) and
(\ref{LKord}).

In his original paper \cite{Loewner} L\"owner dealt with functions
that map $U$ onto domains each of which was obtained by slitting
$\mathbb C$ along a unique Jordan curve. This led him to a subclass
$S'$ of the class $S$ of one-slit maps which is dense in $S$.  Each
function $f\in S'$ is obtained as the limit (\ref{limit}) where
$w(z,t)$ is a solution to the Cauchy problem for the L\"owner
equation
\begin{equation}
\frac{dw}{dt}=-w\frac{e^{iu(t)}+w}{e^{iu(t)}-w},\label{Lord}
\end{equation}
with the initial condition $w(z,0)=z$ where $u(t)$ is a continuous
control function. The corresponding L\"owner equation in partial
derivatives is just equation (\ref{LK}) with the function
$p(\zeta,t)$ given by (\ref{yadro}). {\bf The solution to this
equation is univalent all the time if and only if the initial
condition is given as the limit (\ref{limit}) for the same $p$ given
by (\ref{yadro})}. It maps the unit disk onto a growing family of
one-slit domains. The class of such domains lies dense in the space
of all simply connected domains in the sense of Carath\'eodory's
kernel convergence.

We proceed with generalized L\"owner chains of domains with several
slits. Let $\Omega(t)$, $0\in \Omega(t)$, $t\in [0,T)$, be a
subordination chain of simply connected domains each of which is
obtained by slitting $\mathbb C$ along a finite number of Jordan
curves with $m+1$ distinct endpoints. The $m+1$-th endpoint is at
infinity. The one-parameter family $f(\zeta,t)$ stands for the
corresponding subordination chain of normalized conformal maps
$f(\zeta,t)=e^t\zeta+b_2(t)\zeta^2+\dots$, $\zeta\in U$, $t\in
[0,T)$. These maps satisfy the equation (\ref{LK}) with the function
$p(\zeta,t)$ given by (\ref{yadro}) where $u(t)$ is a piecewise
continuous control function regarding to $t\in[0,T)$ (see, e.g.,
Goluzin \cite{Goluzin}). However, this representation is not unique
and another approach has been suggested originally by Kufarev and
described in \cite{Aleksandrov}. It was based on the function
$p(\zeta,t)$ in (\ref{LK}) given by the following formula
\begin{equation}
p(\zeta,t)=\sum\limits_{k=1}^m\lambda_k(t)\frac{e^{iu_k(t)}+\zeta}{e^{iu_k(t)}-\zeta},\label{yadro2}
\end{equation}
where $u_k(t)$ are continuous control functions regarding to
$t\in[0,T)$ and $\lambda_k(t)$ are measurable non-negative control
functions, $\sum_{k=1}^m\lambda_k(t)=1$.

 Constructing the characteristic equation (\ref{LKord}) we come to
 the L\"owner-Kufarev equation in ordinary derivatives and the limit
 (\ref{limit}) gives a subclass $S''$ of the class $S$ of multi-slit
 maps which is also dense in $S$. The function $w(z,t)$ in
 (\ref{limit}) is a solution to the Cauchy problem for the
 L\"owner-Kufarev
equation
\begin{equation}
\frac{dw}{dt}=-w\sum\limits_{k=1}^m\lambda_k(t)\frac{e^{iu_k(t)}+w}{e^{iu_k(t)}-w},\label{LKord2}
\end{equation}
with the initial condition $w(z,0)=z$ where $u_k(t)$ and
$\lambda_k(t)$ are defined as in (\ref{yadro2}).

Similarly to the one-slit evolution, the equation (\ref{LK}) with
the function $p$ given by (\ref{yadro2}) possesses univalent
solutions all the time if and only if the initial condition is given
as the limit (\ref{limit}) for the same $p$ in (\ref{LKord2}) given
by (\ref{yadro2}).

Although the representation ({\ref{LK}}, \ref{yadro2}) or
(\ref{LKord2}) is not unique neither, it allows to elaborate a
unique one. Given a subordination chain $\Omega(t)$ of multi-slit
domains there exists a unique set of positive numbers
$\{\lambda_k\}_{k=1}^m$, $\sum_{k=1}^m\lambda_k=1$, and a unique set
of continuous control functions $\{u_k(t)\}_{k=1}^m$, such that
$\Omega(t)=f(U,t)$, where $f$ is a solution to the equation
(\ref{LK}) with the function $p$ given by (\ref{yadro2}), and
$\lambda_k(t)\equiv \lambda_k$. This statement follows from the
results of \cite{Prokhorov}. An analogous uniqueness statement for
the representation (\ref{LKord2}) is true too.

\section{Evolution parameters and hierarchies}

As it was stated in Introduction, one of the examples of integrable
systems of evolution parameters corresponding to one-parameter
families of domains is the problem of Laplacian growth. In its
classical formulation it deals with the smooth family
$\{\Omega(t)\}$, $t\in [0,T)$ of phase domains, i.e., when $\partial
\Omega(t)$ are smooth ($C^{\infty}$) interfaces for each $t$, and
the normal velocity $v_n$ at the boundary continuously depends on
$t$ at any point of $\partial \Omega(t)$. Each $\Omega(t)$ is
supposed to be simply connected in $\mathbb C$ for any $t\in [0,T)$
fixed. Starting from the initial moment $t=0$ the boundary becomes
even analytic smooth for $t\in (0,T)$. Richardson's complex moments
completely describe this evolution, say given an initial phase
domain $\Omega(0)$ there exists a unique infinite set of the moments
for $\Omega(t)$ at each  $t\in (0,T)$ and vice versa. One of the
reasons for a slit dynamics to appear in this process is extending
the solution to exist beyond the blowup time $T$. One way of the
regularization of the problem is to introduce a surface tension
parameter instead of zero boundary condition in the classical
Hele-Shaw formulation (see, e.g., \cite{How}). This can produce a
crack morphology in which a part of the boundary develops a thin
finger penetrating the phase domain. Another proposition (see
\cite{Hohlov}) is to continue smooth dynamics by slit dynamics
beyond the blowup time when a possible cusp is developed at the
boundary. Of course, in this case one relaxes smoothness of the
boundary permitting the corresponding Riemann map to have
singularities at the unit circle.

Another reason to consider slit dynamics is the determinate
evolution, duality, and univalence of the solutions to the L\"owner
and L\"owner-Kufarev equations (in ordinary and partial derivatives)
stated in the preceding section.

In view of the involved Riemann maps the most natural evolution
parameters are the Taylor coefficients of these maps. Of course, the
infinite set of coefficients completely describes the evolution
given by the maps $f:\, U\to \Omega(t)$. However in general, the
differential equations for these coefficients obtained from the
L\"owner or L\"owner-Kufarev equations do not admit any reasonable
Hamiltonian formulation that leads to integration of this
(infinitely dimensional) system.

Instead we propose to construct  hierarchies defined by the
coefficient bodies as an object to study. Let $f\in S$. We construct
the coefficient bodies by setting
\[
V_n=\{(b_2,\dots, b_n):\,
f(\zeta)=\zeta+\sum\limits_{k=2}^{\infty}b_k\zeta^k\in S\},\quad
n\geq 2,
\]
in the $(2n-2)$-dimensional Euclidean space. Every point of the
boundary of $V_n$ is given by a unique function $f\in S$ that maps
the unit disk onto the plane minus piecewise analytic Jordan arcs
forming a tree with a root at infinity having at most $n-1$ tips.
Each $V_n$ defines a class $\{\Omega_0\}$ of initial domains for
multi-slit subordination evolutions $\{\Omega(t)\}$, and the
corresponding systems of differential equations for coefficients
form an integrable system $D_n$ for each $n$, moreover, its dual
system $D_n^*$ obtained by the characteristic equation admits
Hamiltonian formulation. The latter systems form hierarchies
$\{D_n\}_{n=2}^{\infty}$ and $\{D_n^*\}_{n=2}^{\infty}$. This idea
up to some extent will be proved in the next sections.

\section{Coefficient bodies}

By the {\it coefficient problem for univalent functions} we mean the
problem of precise finding the regions $V_n$ defined above. These
compact sets have been investigated by a great number of authors,
but the most remarkable source is a famous monograph \cite{SS}
written by Schaeffer and Spencer in 1950. Among other contributions
to the coefficient problem we distinct a monograph by Babenko
\cite{Babenko} that contains a good collection of qualitative
results on the coefficient bodies $V_n$. The results concerning the
structure and properties of $V_n$ include

\begin{itemize}

\item[(i)] $V_n$ is homeomorphic to a $(2n-2)$-dimensional ball and its
boundary $\partial V_n$ is homeomorphic to a $(2n-3)$-dimensional
sphere;

\item[(ii)] every point $x\in \partial V_n$ corresponds to exactly one function $f\in
S$ which will be called a {\it boundary function} for $V_n$;

\item[(iii)] with the exception for a set of smaller dimension,
at every point $x\in \partial V_n$ there exists a normal vector
satisfying the Lipschitz condition;

\item[(iv)] there exists a connected open set $X_1$ on $\partial V_n$,
such that the boundary $\partial V_n$ is an analytic hypersurface at
every point of $X_1$. The points of $\partial V_n$ corresponding to
the functions that give the extremum to a linear functional belong
to the closure of $X_1$.

\end{itemize}

It is worth to note that all boundary functions have a similar
structure. They map the unit disk $U$ onto the complex plane
$\mathbb C$ minus piecewise analytic Jordan arcs forming a tree with
a root at infinity and having at most $n-1$ tips. This assertion
underlines the importance of multi-slit maps in the coefficient
problem for univalent functions.

L\"owner's approach is based on the following idea: a function
mapping $U$ onto the complement of a single slit admits univalent
dynamics for $t\in [0,\infty)$ represented by the L\"owner equation
(\ref{LK}-\ref{yadro}) forming a subordination chain
$f(\zeta,t)=\sum_{n=1}^{\infty} b_n(t)\zeta^n$. Let us deduce a
system of differential equations for the coefficients of
$f(\zeta,t)$ substituting the expansion of $f(\zeta,t)$ into
(\ref{LK}). This gives
\[
\dot{b}_k=kb_k+2\sum_{j=1}^{k-1}jb_je^{-i(k-j)u},\quad b_k(0)=a_k,
\quad t\geq 0,
\]
$k=1,2,\dots,$ where $f(\zeta,0)=f_0(\zeta)=\sum_{n=1}^{\infty}
a_n\zeta^n$. In particular, $b_1(t)=e^t$.

Going to the L\"owner equation in characteristics (\ref{Lord}) we
write $w(z,t)=\sum_{n=1}^{\infty} a_n(t)\zeta^n$ and substitute this
function in (\ref{Lord}). To formulate the result we introduce the
following notations
\[
a(t)=\left(
\begin{array}{c}
a_1(t)\\
\cdot\\
\cdot\\
\cdot\\
a_n(t)
\end{array} \right),\quad
A(t)=\left(
\begin{array}{lllll}
0 & 0 & \dots & 0 & 0\\
a_1(t) & 0 & \dots & 0 & 0\\
a_2(t) & a_1(t) & \dots & 0 & 0\\
\dots & \dots & \dots & \dots & \dots\\
a_{n-1}(t) & a_{n-2}(t) & \dots & a_{1}(t) & 0
\end{array} \right).
\]
Then the differential equation for $a(t)$ is of the form
\begin{equation}
\dot{a}=-a-2\sum_{s=1}^{n-1}e^{-isu}A^sa,\label{coeff}
\end{equation}
\[
a(0)\equiv a^0=\left(
\begin{array}{c}
1\\
0\\
\cdot\\
\cdot\\
\cdot\\
0
\end{array} \right).
\]
In particular, $a_1(t)=e^{-t}$ and $\lim_{t\to\infty}e^ta_k(t)=a_k$.

\section{Hamiltonian formulation for the coefficient system. Integrability}

\subsection{Hamiltonian dynamics and integrability}

Let us recall briefly the Hamiltonian and symplectic definitions and
concepts that will be used in the sequel. There exists a vast amount
of modern literature dedicated to different approaches to and
definitions of {\it intagrable systems} (see, e.g., \cite{Arnold},
\cite{Babelon}, \cite{Bolsinov}, \cite{Zakharov}).

The classical definition of a {\it completely integrable system} in
the sense of Liouville applies to a Hamiltonian system. If we can
find  independent conserved integrals which are pairwise involutory
(vanishing Poisson bracket), this system is completely integrable
(see e.g., \cite{Arnold}, \cite{Babelon}, \cite{Bolsinov}). That is
each first integral allows us to reduce the order of the system not
just by one, but by two. We formulate this definition in a slightly
adopted form as follows.

A dynamical system in $\mathbb C^{2n}$ is called {\it Hamiltonian}
if it is of the form
\begin{equation}
\dot{x}=\nabla_s H(x),\label{Hamilton1}
\end{equation}
where $\nabla_s$ denotes the {\it symplectic gradient} given by
\[
\nabla_s=\left(\frac{\partial}{\partial
\bar{x}_{n+1}},\dots,\frac{\partial}{\partial
\bar{x}_{2n}},-\frac{\partial}{\partial
{x}_1},\dots,-\frac{\partial}{\partial {x}_n}\right).
\]
The function $H$ in (\ref{Hamilton1}) is called the {\it Hamiltonian
function} of the system. It is convenient to redefine the
coordinates $(x_{n+1},\dots, x_{2n})=(\psi_{1},\dots, \psi_{n})$,
and rewrite the system (\ref{Hamilton1}) as
\begin{equation}
\dot{x}_k=\frac{\partial H}{\partial \overline{\psi}_k},\quad
\dot{\overline{\psi}}_k=-\frac{\partial H}{\partial x_k},\quad
k=1,2\dots,n.\label{Hamilton2}
\end{equation}
The system has $n$ degrees of freedom. The two-form
$\omega=\sum_{k=1}^n dx\wedge d\bar{\psi}$ admits the standard
Poisson bracket $\{\cdot,\cdot\}$
\[
\{f,g\}=\sum\limits_{k=1}^{n}\left(\frac{\partial f}{\partial x_k}
\frac{\partial g}{\partial \overline{\psi}_k}-\frac{\partial
f}{\partial \overline{\psi}_k} \frac{\partial g}{\partial
x_k}\right)
\]
associated with $\omega$. The symplectic pair $(\mathbb C^{2n},
\omega)$ defines the Poisson manifold $(\mathbb C^{2n},
\{\cdot,\cdot\})$. These notations may be generalized for a
symplectic manifold and a Hamiltonian dynamical system on it.

The system (\ref{Hamilton2}) may be rewritten  as
\begin{equation}
\dot{x}_k=\{x_k, H\},\quad
\dot{\overline{\psi}}_k=\{\overline{\psi}_k, H\},\quad
k=1,2\dots,n,\label{Hamilton3}
\end{equation}
and the {\it first integrals} $\Phi$ of the system are characterized
by
\begin{equation}
\{\Phi, H\}=0.\label{Hamilton4}
\end{equation}
In particular, $\{H,H\}=0$, and the Hamiltonian function $H$ is an
integral of the system (\ref{Hamilton1}). If the system
(\ref{Hamilton3}) has $n$ functionally independent integrals
$\Phi_1,\dots,\Phi_n$, which are pairwise involutory
$\{\Phi_k,\Phi_j\}=0$, $k,j=1,\dots,n$, then it is called {\it
completely integrable} in the sense of Liouville. The function $H$
is included in the set of the first integrals. The classical theorem
of Liouville and Arnold \cite{Arnold} gives a complete description
of the motion generated by the completely integrable system
(\ref{Hamilton3}). It states that such a system admits action-angle
coordinates around a connected regular compact invariant manifold.
One can work with real Poisson manifolds instead, making use of the
{\it real Hamiltonian function} $2\R H$ keeping all other formulas
changeless.

If the Hamiltonian system admits only $1\leq k<n$ independent
involutory integrals, then it is called {\it partially integrable}.
The case $k=1$ is known as the Poincar\'e--Lyapunov theorem which
states that a periodic orbit of an autonomous Hamiltonian system can
be included in a one-parameter family of such orbits under a
nondegeneracy assumption. A bridge between these two extremal cases
$k=1$ and $k=n$ has been proposed by Nekhoroshev \cite{Nekhoroshev}
and proved later in \cite{Bambusi}, \cite{Fiorani}, \cite{Gaeta}.
The result states the existence of $k$-parameter families of tori
under suitable nondegeneracy conditions.

\subsection{Coefficient system}

We will show that the system (\ref{coeff}) becomes an integrable
system when treated as a description of the boundary hypersurface
$\partial V_n$. Solutions $a(t)$ to (\ref{coeff}) for different
control function $u(t)$ (piecewise continuous, in general) being
multiplied by the factor $e^t$ represent  all points of $\partial
V_n$ as $t\to\infty$. The trajectories $e^ta(t)$, $0\leq t <
\infty$, fill $V_n$ so that every point of $V_n$ belongs to a
certain trajectory $e^ta(t)$. The endpoints of these trajectories
can be interior or else boundary points of $V_n$. In this way, we
set $V_n$ as the closure of the reachable set for the control system
(\ref{coeff}).

According to property (ii) of $V_n$ given in the previous section,
every point $x\in \partial V_n$ is attained by exactly one
trajectory  $e^ta(t)$ which is determined by a choice of the
piecewise continuous control function $u(t)$. The function $f\in S$
corresponding to $x$  is a multi-slit map of $U$. If the boundary
tree of $f$ has only one tip, then there is a unique continuous
control function $u(t)$ in $t\in [0,\infty)$ that corresponds to
$f$. The case of multi-slit maps we will consider in the next
section.

To reach a boundary point $x\in \partial V_n$ corresponding to a
one-slit map, the trajectory $e^ta(t)$ has to obey extremal
properties, i.e., to be an {\it optimal trajectory}. The continuous
control function $u(t)$ must be optimal, and hence, it satisfies a
necessary condition of optimality. {\it The Maximum Principle} is a
powerful tool to be used that provides a joint interpretation of two
classical necessary variational conditions: the Euler equations and
the Weierstrass inequalities (see, e.g., \cite{Pontryagin}).

To realize the maximum principle we consider an adjoint vector
\[
\psi(t)=\left(
\begin{array}{c}
\psi_1(t)\\
\cdot\\
\cdot\\
\cdot\\
\psi_n(t)
\end{array} \right),
\]
with complex valued coordinates $\psi_1,\dots,\psi_n$, and the
real Hamiltonian function
\[
H(a,{\psi},u)=2\R\left[\left(-a-2\sum\limits_{s=1}^{n-1}e^{-isu(t)}A^sa\right)^T\bar{\psi}\right],
\]
where $\bar{\psi}$ means the vector with complex conjugate
coordinates. To come to the Hamiltonian formulation for the
coefficient system we require that $\bar{\psi}$ satisfies the
adjoint system of differential equations
\begin{equation}
\frac{d\bar{\psi}}{dt}=-\frac{\partial H}{\partial a},\quad 0\leq
t<\infty.\label{psi}
\end{equation}
Taking into account (\ref{coeff}) we rewrite (\ref{psi}) as
\begin{equation}
\frac{d\bar{\psi}}{dt}=\left(E+2\sum\limits_{s=1}^{n-1}e^{-isu(t)}(s+1)(A^T)^s\right)\bar{\psi},\label{psi2}
\end{equation}
where $E$ is the unit matrix.

The maximum principle states that any optimal control function
$u^*(t)$ possesses a maximizing property for the Hamiltonian
function along the corresponding trajectory, i.e.,
\begin{equation}
\max\limits_{u}H(a^*(t),\psi^*(t),u)=H(a^*(t),\bar{\psi}^*(t),u^*),
\quad t\geq 0, \label{max}
\end{equation}
where $a^*$ and $\psi^*$ are solutions to the system (\ref{coeff},
\ref{psi2}) with $u=u^*(t)$.

The maximum principle (\ref{max}) yields that
\begin{equation}
\frac{\partial H(a^*(t),\psi^*(t),u)}{\partial
u}\bigg|_{u=u^*(t)}=0.\label{max1}
\end{equation}
Evidently, (\ref{coeff}), (\ref{psi}), and (\ref{max1}) imply that
\begin{equation}
\frac{d\, H(a^*(t),\psi^*(t),u^*(t))}{dt}=0,\label{ham}
\end{equation}
for an optimal differentiable control function $u^*(t)$.

The Hamiltonian formalism for system (\ref{coeff}, \ref{psi2}) will
lead to integrability. First, we show how $\psi$ can be expressed in
terms of the phase variable $a$. In the following theorem causing no
confusion we will use ${^{\rm t}}(\dots)$ to denote the matrix transposition.

\begin{theorem}\label{th1}
Let $a(t)$ and $\psi(t)$, ${\psi}(T)={^{\rm t}}(v_1,\dots, v_n)$ obey the
system (\ref{coeff}, \ref{psi}), $T\geq 0$. Then
$\bar{\psi}_k(t)={c}_{n-k+1}$, $k=1,\dots, n$, where $c_1,\dots,
c_n$ are the Taylor coefficients of the expansion
\[
\frac{(\bar{v}_nz+\dots+\bar{v}_1z^n)w'(z,
T)}{w'(z,t)}=\sum\limits_{k=1}^{\infty}c_k(t)z^k.
\]
\end{theorem}
\begin{proof}
Let $w(z,t)=\sum_{k=1}^{\infty}a_k(t)z^k$ be a solution to the
L\"owner differential equation (\ref{Lord}). Differentiating
(\ref{Lord}) with respect to $z$  we immediately have
\begin{equation}
\frac{d}{dt}\left(\frac{z}{w'(z,t)}\right)=\frac{z}{w'(z,t)}
\left(\frac{e^{iu}+w}{e^{iu}-w}+\frac{2e^{iu}w}{(e^{iu}-w)^2}\right).\label{2}
\end{equation}
Considering the expansion
\[
\frac{z}{w'(z,t)}=\sum\limits_{k=1}^{\infty}q_k(t)z^k,
\]
we obtain
\begin{equation}
\frac{d\,q(t)}{dt}=\left(E+2\sum\limits_{s=1}^{n-1}e^{-isu(t)}(s+1)A^s\right)q,\label{3}
\end{equation}
where $q(t)=(q_1(t),\dots, q_n(t))^T$. We observe that system
(\ref{3}) differs from system (\ref{psi2}) only by the transposition
sign.

In order to satisfy the condition $q(T)=(\bar{v}_1,\dots,
\bar{v}_n)^T$ we denote by
\[
g(z,t)=\frac{(\bar{v}_nz+\dots+\bar{v}_1z^n)w'(z,
T)}{w'(z,t)}=\sum\limits_{k=1}^{\infty}c_k(t)z^k,
\]
and observe that $g(z,t)$ obeys the same equation (\ref{2}) where
${z}/{w'(z,t)}$ is substituted by $g(z,t)$. Hence,
\[
c(t)=\left(
\begin{array}{c}
c_1(t)\\
\cdot\\
\cdot\\
\cdot\\
c_n(t)
\end{array} \right)
\]
obeys the system (\ref{3}) substituting $q(t)$ by $c(t)$. It is
easily seen that
\[
c(T)=\left(
\begin{array}{c}
\bar{v}_n\\
\cdot\\
\cdot\\
\cdot\\
\bar{v}_1
\end{array} \right).
\]
The difference in the transposition sign implies that
$\bar{\psi}_k=c_{n-k+1}$, $k=1,\dots, n$. This completes the proof.
\end{proof}

Putting $T=0$ in this theorem we come to the following corollary.

\begin{corollary}\label{cor}
Let $a(t)$ and $\bar{\psi}(t)$, ${\psi}(0)=({v}_1,\dots, v_n)^T$
obey the system (\ref{coeff}, \ref{psi}), $t\geq 0$. Then
$\bar{\psi}_k(t)={c}_{n-k+1}$, $k=1,\dots,n$, where $c_1,\dots, c_n$
are the Taylor coefficients of the expansion
\[
\frac{(\bar{v}_nz+\dots+\bar{v}_1z^n)}{w'(z,t)}=\sum\limits_{k=1}^{\infty}c_k(t)z^k.
\]
\end{corollary}

\begin{remark}
Since $\I a_1(t)=0$, the Hamiltonian function $H(a,\psi, u)$ does
not depend on $\I \psi_1$. Therefore, without loss of generality, we
can put ${\psi}_1(0)=v_1$ to be real. This assumption leave $a$,
$\psi_2,\dots, \psi_n$ changeless.
\end{remark}

Now we are able to apply the maximum principle. An optimal
continuous control function  satisfies the maximizing property
(\ref{max}) and obeys equation (\ref{max1}). The Hamiltonian
$H(a,\psi,u)$ is a trigonometric polynomial with respect to $u$ of
degree $n-1$ if $\psi_n\neq 0$. Let $\psi(0)=(v_1,\dots,v_n)^T$.
Note that ${\psi}_n(t)=e^t$. We assume that $v_n\neq 0$, and
diminish $n$ to $n-1$, otherwise. At $t=0$, we have
\[
H(a^0,(v_1,\dots,v_n)^T, u)=-v_1e^{-t}-2\sum\limits_{s=1}^{n-1}(\R
v_{s+1}\cos(su)+\I v_{s+1}\sin(su)).
\]
For the optimal $u^*$,
\[
H_u(a^0,(v_1,\dots,v_n)^T, u^*)=2\sum\limits_{s=1}^{n-1}(s\,\R
v_{s+1}\sin(su^*)-s\,\I v_{s+1}\cos(su^*))=0.
\]
Suppose that $v_2,\dots, v_n$ are such that
\[
H_{uu}(a^0,(v_1,\dots,v_n)^T, u^*)=2\sum\limits_{s=1}^{n-1}(s^2\,\R
v_{s+1}\cos(su^*)+s^2\,\I v_{s+1}\sin(su^*))\neq 0,
\]
for the optimal $u^*$. As we see, the term with $v_1$ does not
influence the process of maximizing, and we will assume it zero
later in course. The inequality holds for all $(v_2,\dots, v_n)\in
\mathbb R^{2n-2}$ except for a set $A$ of dimension at most $2n-4$.
Indeed, due to the maximizing property of the optimal control
function, this condition breaks down if
\[
H_{uu}(a^0,(v_1,\dots,v_n)^T, u^*)=H_{uuu}(a^0,(v_1,\dots,v_n)^T,
u^*)=0.
\]
Thus, the set $A$ is determined by two linearly independent
equations
\[
\sum\limits_{s=1}^{n-1}(s^2\,\R v_{s+1}\cos(su^*)+s^2\,\I
v_{s+1}\sin(su^*))=0,
\]
\[
\sum\limits_{s=1}^{n-1}(s^3\,\R v_{s+1}\sin(su^*)-s^3\,\I
v_{s+1}\cos(su^*))=0,
\]
with the fixed optimal $u^*$, varying $v_2,\dots,v_n$. Therefore,
$A$ is a linear space of dimension $2n-4$ in $\mathbb R^{2n-2}$.

\begin{definition}
We say that a vector $(v_2,\dots,v_n)$ satisfies the regularity
condition on $[0,T]$ if
\[
H_{uu}(a^*(t), \psi^*(t), u^*)\neq 0, \quad t\in [0,T],
\]
for the optimal $u^*$ and $\psi^*(0)=(v_1,\dots,v_n)^T$ along the
optimal trajectory $(e^ta^*(t),\psi^*(t))$ corresponding to $u^*$.
\end{definition}

Evidently, if $(v_2,\dots,v_n)\in \mathbb R^{2n-2}\setminus A$, then
$(v_2,\dots,v_n)$ satisfies the regularity condition on $[0,T]$ for
$T>0$ and small enough. Let us denote by $Y(T)$ the set of all
$(v_2,\dots,v_n)$ satisfying the regularity condition on $[0,T]$.

Let $(v_2,\dots,v_n)\in Y(T)$. Then
\begin{equation}
H_{uu}(a,\psi, u)\neq 0, \quad t\in [0,T],\label{aa}
\end{equation}
for $(a,\psi, u)$ from a neighborhood of $(a^*,\psi^*, u^*)$ and
this neighborhood depends on $t$. The equation
\[
H_u(a,\psi, u)=0,
\]
together with the regularity condition on $[0,T]$ determines an
analytic implicit function $u=u(a,\psi)$ in a neighborhood of
$(a^*,\psi^*)$, and $u(a,\psi)$ lies in a neighborhood of $u^*\equiv
u(a^*,\psi^*)$.

We substitute $u=u(a,\psi)$ in the phase system (\ref{coeff}) and
the adjoint system (\ref{psi}) and solve the Cauchy problem with the
initial data $a(0)=a^0$, $\psi(0)=({v}_1,\dots,{v}_n)^T$. Since
$u=u^*$ along $(a^*,\psi^*)$, the Hamiltonian $H(a,\psi,u)$
satisfies the maximum principle. Note that the neighborhood of
$(a^*(t), \psi^*(t))$ can be determined by a neighborhood of the
corresponding vector $(v_2,\dots,v_n)=({\psi}_2(0),\dots,
{\psi}_n(0))$. According to Corollary \ref{cor} the function
$\psi(t)$ can be represented as a function of the phase variable $a$
and of the initial conditions $(v_1,\dots,v_n)$, say
$\psi(t)=\varphi(a(t),v_1,\dots,v_n)$.

\begin{theorem}
Let $(v_2,\dots,v_n)$ satisfy the regularity condition  on $[0,T]$.
Then the system (\ref{coeff}, \ref{psi2}) with
$u=u(a,\psi)=u(a,\varphi(a,0,v_2,\dots,v_n))$ is partially Liouville
integrable. Moreover, the first integrals form a contact structure.
\end{theorem}
\begin{proof}
We substitute $\psi(t)=\varphi(a(t),v_1,\dots,v_n)$ in the
Hamiltonian and obtain
\[
H(a(t),\psi(t),u)=H(a(t),\varphi(a(t),v_1,\dots,v_n),
u)=\mathcal{H}(a(t),v_1,\dots,v_n, u).
\]
To each point $(v_1,\dots,v_n)$ from a neighborhood of a vector
satisfying the regularity condition, there corresponds an optimal
continuous control function. The maximum principle implies that
\[
\mathcal{H}_u(a(t),v_1,\dots,v_n,
u)\bigg|_{u=u(a,\varphi(a,v_1,\dots,v_n))}=0,
\]
that, together with (\ref{ham}), gives
\begin{equation}
\mathcal{H}(a(t),v_1,\dots,v_n,
u(a,\varphi(a,v_1,\dots,v_n)))=const.\label{const}
\end{equation}

In order to prove the partial integrability we will find the first
complex integrals $\Phi_1,\dots \Phi_n$. We already have one real
integral (\ref{const}).

By the corollary from Theorem \ref{th1} we deduce that
\begin{equation}
\sum\limits_{k=1}^n\bar{v}_{n-k+1} z^k=w'(z,t)\sum\limits_{k=1}^n\bar{\psi}_{n-k+1} z^k+w'(z,t)\sum\limits_{k=n+1}^{\infty}c_k z^k.\label{firstint1}
\end{equation}
We denote by  ${^{\rm t}}(\Phi_1,\dots, \Phi_n)$ the vector of the
first integrals of the Hamiltonian system (\ref{coeff}, \ref{psi2})
given by
\[
\left(
\begin{array}{c}
\Phi_1\\
\Phi_2\\
\Phi_3\\
\dots\\
\Phi_n
\end{array} \right)=\left(
\begin{array}{rrrrr}
a_1&2a_2 & \dots & (n-1)a_{n-1} & na_n\\
0 & a_1 & \dots &  (n-2)a_{n-2} & (n-1)a_{n-1}\\
0 & 0 & \dots & (n-3)a_{n-3} & (n-2)a_{n-2}\\
\dots & \dots & \dots & \dots & \dots\\
0 & 0 & \dots & 0 & a_1
\end{array} \right)\left(
\begin{array}{c}
\bar{\psi}_1\\
\bar{\psi}_2\\
\bar{\psi}_3\\
\dots\\
\bar{\psi}_n
\end{array} \right),
\]
 with the
control $u$ given in the statement of the theorem. Indeed, the
equality (\ref{firstint1}) implies that $\Phi_k=\bar{v_k}$ are
constants for all $t$ and $k=1,\dots,n$. The first integral
(\ref{const}) allows us to conclude that $\{\Phi_k,\mathcal{H}\}=0$.
One checks that $\{\Phi_s,\Phi_k\}=- s\Phi_{s+k-1}$ for all $1\leq
s<k$ and $s+k-1\leq n$. Otherwise, $\{\Phi_s,\Phi_k\}=0$ for
$s+k-1>n$. This implies that
\begin{itemize}
\item $[n/2]$ first integrals $\{\Phi_{[n/2]+1},\dots, \Phi_n\}$ are
pairwise involutory;
\item the integrals $\{\Phi_1,\dots,\Phi_{[n/2]}\}$ are not pairwise
involutory but their Poisson brackets give all the rest of
integrals. This structure is said to be contact (which is not
uncoupled because, for example,
$\{\Phi_1,\dots,\Phi_{n}\}=-\Phi_n\neq 0$).
\end{itemize}
This completes the proof.
\end{proof}

\begin{remark}
We observe that
\[
H(a_0,(v_1,\dots,v_n)^T, u^*)=-v_1-2\sum\limits_{s=1}^{n-1}(\R
v_{s+1}\cos(su^*)+\I v_{s+1}\sin(su^*))
\]
at $t=0$, which is the constant in the right-hand side of
(\ref{const}). It contains an additive term $(-v_1)$. The optimal
control function $u^*$ does not depend on $v_1$. Hence, we can put
$v_1=0$ without loss of generality. This assumption does not
influence $a, \psi_2,\dots, \psi_n$, and defines uniquely the
constant in the right-hand side of (\ref{const}). Thus, the first
integral (\ref{const}) can be rewritten as
\begin{equation}
\mathcal{H}(a(t),v_2,\dots,v_n,
u(a,\varphi(a,v_2,\dots,v_n)))=const. \label{const2}
\end{equation}
\end{remark}

The integral (\ref{const2}) gives a local parametric description of
the boundary of the coefficient body for bounded univalent functions
from $S$ (such that $f\in S$, $|f(z)|<M$, $1< M\leq e^T$). The
points of this part of the boundary correspond to the functions that
map $U$ onto the disk $|w|<M$ slit along an analytic curve with
exactly one tip at its interior point. The boundary hypersurface is
homeomorphic to an open set on the $2n-3$-dimensional sphere in
$\mathbb R^{2n-2}$ and must be parameterized by $2n-3$ free
parameters. We should select $2n-3$ independent parameters among
$2n-2$ variables $v_2,\dots,v_n$. Observe that the Hamiltonian
$H(a,\psi,u)$ and the solution $\psi$ to the adjoint system
(\ref{psi}) are linear with respect to $\psi$. The implicit function
$u=u(a,\psi)$ is invariant upon multiplying $\psi$ by a positive
number. Therefore, both $H$ and $\psi$ may be determined up to a
positive multiplier without loss of generality. Putting, e.g.,
$|v_n|=1$ we reduce the number of free parameters in (\ref{const2})
to $2n-3$ that gives a local parametric representation of the
boundary.

The integral (\ref{const}) is continued from $[0,T]$ to a bigger
interval $[0,T+\varepsilon)$ as long as the regularity conditions
are preserved. Let $T>0$ be the minimal positive number such that
the regularity conditions break up at $T$ for $v_2\dots,v_n$, i.e.,
\[
\mathcal{H}_{uu}(a(T),v_2,\dots,v_n,
u(a(T),\varphi(a(T),v_2,\dots,v_n)))=0,
\]
\[
\mathcal{H}_{uuu}(a(T),v_2,\dots,v_n,
u(a(T),\varphi(a(T),v_2,\dots,v_n)))=0.
\]
Denote by
\[
G_1(t,v_2,\dots,v_n):=\mathcal{H}_{uu}(a(t),v_2,\dots,v_n,
u(a(t),\varphi(a(t),v_2,\dots,v_n))),
\]
\[
G_2(t,v_2,\dots,v_n):=\mathcal{H}_{uuu}(a(t),v_2,\dots,v_n,
u(a(t),\varphi(a(t),v_2,\dots,v_n))).
\]
In these formulas we admit that $a(t)$ is determined by
$v_2,\dots,v_n$ according to the phase system (\ref{coeff}). The
system $G_1(t,v_2,\dots,v_n)=G_2(t,v_2,\dots,v_n)=0$ of linearly
independent equations defines a $(2n-4)$-dimensional manifold in the
$(2n-3)$-dimensional space of free parameters $v_2,\dots,v_n$,
$|v_n|=1$.

Thus, all integrals (\ref{const2}) are continued on $t\in
[0,\infty)$ for all $v_2,\dots,v_n$ except for a
$(2n-4)$-dimensional set. Using continuity of integrals
(\ref{const2}) with respect to $v_2,\dots,v_n$ we conclude that the
relation (\ref{const2}) holds for all $(v_2,\dots,v_n)\in \mathbb
R^{2n-3}$ and for all $t\geq 0$.

\section{Multi-slit evolution}

Now we generalize the results of the preceding section to the
multi-slit evolution. We do this for the maps with the boundary tree
having exactly two finite tips because other cases can be easily
obtained by analogy. The L\"owner-Kufarev equation (\ref{LKord2})
has the form
\[
\frac{dw}{dt}=-w\left(\lambda\frac{e^{iu_1(t)}+w}{e^{iu_1(t)}-w}+(1-\lambda)\frac{e^{iu_2(t)}+w}{e^{iu_2(t)}-w}\right),
\]
where $u_1$ and $u_2$ are continuous control functions and
$\lambda\in (0,1)$. The differential equation for phase variables
(\ref{coeff}) becomes
\begin{equation}
\dot{a}=-a-2\sum_{s=1}^{n-1}(\lambda e^{-isu_1}+(1-\lambda)
e^{-isu_2})A^sa,\label{lcoeff}
\end{equation}
The Hamiltonian is written as
\[
\tilde{H}(a,{\psi},u_1,u_2,\lambda)=\R\left[\left(-a-2\sum\limits_{s=1}^{n-1}(\lambda
e^{-isu_1}+(1-\lambda) e^{-isu_2})A^sa\right)^T\bar{\psi}\right],
\]
and the adjoint system (\ref{psi2}) as
\begin{equation}
\frac{d\bar{\psi}}{dt}=\left(E+2\sum\limits_{s=1}^{n-1}(\lambda
e^{-isu_1}+(1-\lambda)
e^{-isu_2})(s+1)(A^T)^s\right)\bar{\psi}.\label{lpsi2}
\end{equation}

The maximizing condition (\ref{max}) now splits into two bits
\begin{equation}
\max\limits_{u_1,u_2,\lambda}\tilde{H}(a^*(t),\psi^*(t),u_1,u_2,\lambda)=\tilde{H}(a^*(t),\bar{\psi}^*(t),u^*_1,u^*_2,\lambda),
\quad t\geq 0, \label{lmax}
\end{equation}
where $a^*$ and $\psi^*$ are solutions to the system (\ref{lcoeff},
\ref{lpsi2}) with $u_1=u_1^*(t)$,  $u_2=u_2^*(t)$, and moreover,
\begin{equation}
\R\left[\left(\sum\limits_{s=1}^{n-1}(e^{-isu^*_1}-
e^{-isu^*_2})A^sa\right)^T\bar{\psi}\right]=0.\label{add}
\end{equation}
Indeed, if condition (\ref{add}) were not true the value of the
optimal $\lambda$ would be 0 or 1, and the corresponding boundary
function would be a one-slit map that contradicts our supposition.
The maximum principle (\ref{lmax}) implies that
\begin{equation}
\frac{\partial \tilde{H}(a^*(t),\psi^*(t),u_1,u_2,\lambda)}{\partial
u_j}\bigg|_{u_j=u_j^*(t)}=0, \quad j=1,2.\label{lmax1}
\end{equation}
Evidently, (\ref{lcoeff}), (\ref{lpsi2}), and (\ref{lmax1}) imply
that
\begin{equation}
\frac{d\,
\tilde{H}(a^*(t),\psi^*(t),u^*_1,u^*_2,\lambda)}{dt}=0,\label{lham}
\end{equation}
for  optimal differentiable control functions $u_1^*(t)$ and
$u_2^*(t)$.

Obviously, Theorem \ref{th1} and the corollary thereafter still true
in this case and $\psi$ can be expressed in terms of the phase
variable $a$. Continuous optimal control functions $u_1^*$, $u_2^*$
satisfy the maximizing property (\ref{lmax}) and obey equations
(\ref{lmax1}). The Hamiltonian $\tilde{H}(a,\psi,u_1,u_2,\lambda)$
is a trigonometric polynomial with respect to $u_1$ and $u_2$ of
degree $n-1$ if $\psi_n\neq 0$. Let again
$\psi(0)=(v_1,\dots,v_n)^T$, and we assume that $v_n\neq 0$, and
diminish $n$ to $n-1$, otherwise. We have ${\psi}_n(t)=e^t$. An
additional parameter is $\lambda\in (0,1)$, which is constant. At
$t=0$, we have
\[
\tilde{H}(a^0,(v_1,\dots,v_n)^T,u_1,u_2,\lambda)=\lambda
H(a^0,(v_1,\dots,v_n)^T,u_1)+(1-\lambda)H(a^0,(v_1,\dots,v_n)^T,u_2).
\]
The vector of parameters is $(\lambda, v_1,\dots,v_n)$, and we put
$v_1=0$ by the same reason as in the preceding section. Let us
consider $(\lambda,v_2,\dots,v_n)\in \mathbb R^{2n-1}$ (or better
$(0,1)\times \mathbb R^{2n-2}$). For the optimal $u_1^*$ and $u_2^*$
\[
\tilde{H}_{u_1}(a^0,(v_1,\dots,v_n)^T,u^*_1,u^*_2,\lambda)=\tilde{H}_{u_2}(a^0,(v_1,\dots,v_n)^T,u^*_1,u^*_2,\lambda)=0.
\]
Suppose that $\lambda,v_2,\dots,v_n$ are such that
\[
\tilde{H}_{u_1u_1}(a^0,(v_1,\dots,v_n)^T,u^*_1,u^*_2,\lambda),
\quad\mbox{or}\quad
\tilde{H}_{u_2u_2}(a^0,(v_1,\dots,v_n)^T,u^*_1,u^*_2,\lambda),
\]
do not vanish. This holds for all $(\lambda,v_2,\dots,v_n)\in
B\subset (0,1)\times \mathbb R^{2n-2}$ except for a set $\tilde{A}$
of dimension at most $2n-4$. The linear space $B$ is obtained by
restriction (\ref{add}) which,  being rewritten at $t=0$ for the
optimal control functions, is equivalent to
\[
H(a^0,(v_1,\dots,v_n)^T,u^*_1)-H(a^0,(v_1,\dots,v_n)^T,u^*_2)=0.
\]
So, the dimension of $B$ is $2n-2$. The set $\tilde{A}$ is
determined by two linearly independent equations
\[
\tilde{H}_{u_1u_1}(a^0,(v_1,\dots,v_n)^T,u^*_1,u^*_2,\lambda)=
\tilde{H}_{u_1u_1u_1}(a^0,(v_1,\dots,v_n)^T,u^*_1,u^*_2,\lambda)=0,
\]
or
\[
\tilde{H}_{u_2u_2}(a^0,(v_1,\dots,v_n)^T,u^*_1,u^*_2,\lambda)=
\tilde{H}_{u_2u_2u_2}(a^0,(v_1,\dots,v_n)^T,u^*_1,u^*_2,\lambda)=0,
\]
with the fixed optimal $u_1^*$, $u_2^*$, varying
$(\lambda,v_2,\dots,v_n)$ in $B$. Therefore, $\tilde{A}$ is a linear
space of dimension $2n-4$ in $B\subset \mathbb R^{2n-2}$.

\begin{definition}
We say that a vector $(\lambda, v_2,\dots,v_n)$ satisfies the
regularity condition on $[0,T]$ for multi-slit evolution if
\[
\tilde{H}_{u_1u_1}(a^*,\psi^*,u^*_1,u^*_2,\lambda)\neq 0, \quad
\mbox{or \ }\tilde{H}_{u_2u_2}(a^*,\psi^*,u^*_1,u^*_2,\lambda)\neq
0, \quad t\in [0,T],
\]
for the optimal $u_1^*$, $u_2^*$ and $\psi^*(0)=(v_1,\dots,v_n)^T$
along the optimal trajectory $(e^ta^*(t),\psi^*(t))$ corresponding
to $u_1^*$, $u_2^*$.
\end{definition}

If  $(\lambda, v_2,\dots,v_n)\in B\setminus \tilde{A}$, then
$(\lambda, v_2,\dots,v_n)$ satisfies the regularity condition on
$[0,T]$ for $T>0$ and small enough. Let us denote by $\tilde{Y}(T)$
the set of all $(\lambda, v_2,\dots,v_n)$ satisfying the regularity
condition on $[0,T]$.

Let $(v_2,\dots,v_n)\in \tilde{Y}(T)$. Then
\[
\tilde{H}_{u_1u_1}(a,\psi,u_1,u_2,\lambda)\neq 0, \quad \mbox{or \
}\tilde{H}_{u_2u_2}(a,\psi,u_1,u_2,\lambda)\neq 0, \quad t\in [0,T],
\]
for $(a,\psi, u_1,u_2,\lambda)$ from a neighborhood of $(a^*,\psi^*,
u^*_1,u^*_2,\lambda)$ and this neighborhood depends on $t$. The
equations
\[
H_{u_1}(a,\psi,u_1,u_2,\lambda)=H_{u_2}(a,\psi,u_1,u_2,\lambda)=0,
\]
together with the regularity condition on $[0,T]$ determine analytic
implicit functions $u_1=u_1(a,\psi,\lambda)$
$u_2=u_2(a,\psi,\lambda)$ in a neighborhood of
$(a^*,\psi^*,\lambda)$, and each of $u_1=u_1(a,\psi,\lambda)$
$u_2=u_2(a,\psi,\lambda)$ lies in a neighborhood of $u_1^*\equiv
u_1(a^*,\psi^*,\lambda)$, $u^*_2\equiv u_2(a,\psi,\lambda)$
respectively.

We substitute $u_1=u_1(a,\psi,\lambda)$, $u_2=u_2(a,\psi,\lambda)$
in the phase system (\ref{lcoeff}) and the adjoint system
(\ref{lpsi2}) and solve the Cauchy problem with the initial data
$a(0)=a^0$, $\psi(0)=({v}_1,\dots,{v}_n)^T$, $(\lambda,
v_2,\dots,v_n)\in B$.  According to Corollary \ref{cor} the function
$\psi(t)$ can be represented as a function of the phase variable $a$
and the initial conditions $(v_1,\dots,v_n)$, say
$\psi(t)=\varphi(a(t),v_1,\dots,v_n,\lambda)$. As in the preceding
section, both $\tilde{H}$ and $\psi$ may be determined up to a
positive multiplier without loss of generality. Putting, e.g.,
$|v_n|=1$ we reduce the number of free parameters to $2n-3$ that
gives a local parametric representation of the boundary. Repeating
all further steps of preceding section we prove the following
theorem.

\begin{theorem}
Let $(\lambda, v_2,\dots,v_n)$ satisfy the regularity condition  on
$[0,T]$. Then the system (\ref{lcoeff}) with
$$u_1=u_1(a,\psi,\lambda)=u_1(a,\varphi(a(t),v_1,\dots,v_n,\lambda),\lambda),$$
$$u_2=u_1(a,\psi,\lambda)=u_2(a,\varphi(a(t),v_1,\dots,v_n,\lambda),\lambda),$$ is partially Liouville integrable.
Moreover, this statement is continued for all $(\lambda,
v_2,\dots,v_n)\in B$ and for all $t\geq 0$.
\end{theorem}

Theorems 2,3 also solve the problem of integrability for the
corresponding coefficient systems in Section 4 substituting the
optimal control functions in the Cauchy problem. One can find exact
solutions in explicit form for $n=3$ in, e.g., \cite{Aleksandrov}
and \cite{Vasiliev}.

 \end{document}